# Kriging based Surrogate Modeling for Fractional Order Control of Microgrids

Indranil Pan and Saptarshi Das

*Abstract*—This paper investigates the use of fractional order (FO) controllers for a microgrid. The microgrid employs various autonomous generation systems like wind turbine generator (WTG), solar photovoltaic (PV), diesel energy generator (DEG) and fuel-cells (FC). Other storage devices like the battery energy storage system (BESS) and the flywheel energy storage system (FESS) are also present in the power network. An FO control strategy is employed and the FO-PID controller parameters are tuned with a global optimization algorithm to meet system performance specifications. A kriging based surrogate modeling technique is employed to alleviate the issue of expensive objective function evaluation for the optimization based controller tuning. Numerical simulations are reported to prove the validity of the proposed methods. The results for both the FO and the integer order (IO) controllers are compared with standard evolutionary optimization techniques and the relative merits and demerits of the kriging based surrogate modeling are discussed. This kind of optimization technique is not only limited to this specific case of microgrid control but also can be ported to other computationally expensive power system optimization problems.

## I. INTRODUCTION

PENETRATION of renewable energy technologies in electrical power systems, in recent years, have increased the system complexity and requires efficient monitoring and control methods to ensure smooth operation of the whole system [1]. The distributed generation (DG) model [2] which uses small capacity generators, typically in the order of 5kW to 10MW, can easily include renewable energy storage systems and other fuel based generators. The DG model offers the advantage of localized generation with consequent reduction in transmission costs and losses. They offer increased reliability and ease of maintenance. Retrofitting other units to the DG is also simple and helps in easier capacity planning and improvement in later stages of operation. Smart grid refers to the integration of these DGs into the grid where there is significant interplay of information and communication technologies for increasing the grid flexibility and system reliability [3]. These smart grids may be composed of different smaller units known as microgrids (MG) [4]. These MGs are comprised of small generation and load units connected at multiple points and can either operate autonomously as an isolated system or can work in a grid connected mode. Since the generating units are small in capacity, their inertia is smaller. This results in severe fluctuations in system parameters like frequency and voltage in cases where the input is stochastic (wind or solar) or there are outages in the generating units. To improve the system stability and performance, energy storage devices like flywheels, batteries and ultra-capacitors are often used [4]. These serve as backup devices and store excess power when the generation is more than the demand and release power to the grid when the demand is more than the generation. This essentially helps in maintaining a steady flow of power irrespective of the generation and load power level fluctuations and consequently keeps the deviations in system frequency at acceptable levels.

For efficient operation of these interconnected generators and energy storage devices, the control of microgrids has received increased attention in recent years [5]. Proper management and control of domestic smart grid technology can be achieved by good prediction, advanced planning and real time control [6]. This helps in a better matching of demand and supply. There are different levels of control schemes in the grid *viz.* the local controls, centralized controls and decentralized controls [5]. Recently intelligent frequency control techniques using particle swarm optimization (PSO) and fuzzy logic have been used in the context of microgrids [7] with encouraging results. Multi-agent system for microgrid control has been investigated in [8]. Genetic algorithms have also been employed for frequency control design in hybrid energy generation/storage system [9].

Application of fractional calculus based control system designs have gained impetus in recent times due to the flexibility and effectiveness that can be gained through such methodologies [10][11]. Merging computational intelligence techniques with fractional order controller designs are also being recently explored [12]. However applicability of FO controllers for electrical power systems is still largely unexplored. A few studies have been done for the application of the fractional order PID (FOPID) controller to the design of the automatic voltage regulator (AVR) and load-frequency control (LFC) in a power system using single objective [13]

I. Pan is with Energy, Environment, Modelling and Minerals (E²M²) Research Section, Department of Earth Science and Engineering, Imperial College London, Exhibition Road, SW7 2AZ, UK (email: i.pan11@imperial.ac.uk).

S. Das is with Communications, Signal Processing and Control (CSPC) Group, School of Electronics and Computer Science, University of Southampton, Southampton SO17 1BJ, UK (e-mail: s.das@soton.ac.uk).

and multi-objective formalisms in frequency domain [14] and in time domain [15] and has been shown to give better results over the traditional PID controller. The FOPID controller has also been applied to the problem of two area load frequency control in a deregulated environment [16]. A fractional calculus based maximum photovoltaic power tracking (MPPT) controller is also proposed for microgrid system in [17].

Most of these optimization based controller design problems involve multiple calls to computationally expensive time domain simulations for power system's dynamics with different guess values of controller parameters which are refined iteratively. This becomes computationally prohibitive and therefore algorithms which can arrive at optimal solutions in less number of iterations are necessary. Meta-models or surrogate modelling methodology for evolutionary algorithms are expedient in such circumstances [18]. These meta-models approximate the fitness landscape of the expensive optimization function and take much less computational time to simulate. Surrogate based modelling methodologies have also been expedient in problems involving dynamic optimization [19], constrained optimization, combinatorial optimization [20] etc. amongst many others. Various techniques can be used for the construction of the surrogate fitness function like radial basis functions [21], neural networks [22], kriging methods [20], response surface models [23] etc. However, surrogate modelling techniques have not been popular for optimization and design of expensive dynamic power system models.

In this paper, a kriging assisted surrogate modelling methodology is outlined and embedded within a global optimization framework for the design of FOPID controllers for microgrid frequency control. The FO differ-integral operators are inherently infinite dimensional linear filters [10], but for practical realization of such systems, band-limited higher order linear system approximations are commonly used [24]. Therefore time domain simulation of such coupled very high order approximated FO systems with several power system components, is computationally expensive. The surrogate modelling methodology is hence suitable in such circumstances to obtain the controller parameters in less number of iterations to pave the path of online tuning this type of FO controllers.

## II. THEORETICAL BACKGROUND

### A. Fractional Calculus Basics

Fractional calculus extends the common notion of integer order integration/differentiation to any arbitrary real number. It can be represented by $_aD_t^\alpha$ where $\alpha \in \Re$ is the order of the differentiation/integration and $a$ and $t$ are the bounds of the operation. There are many definitions of fractional calculus like the Grünwald-Letnikov (GL), Riemann-Liouville (RL) and Caputo definitions [11]. In control system studies, the Caputo definition is mostly used for realizing the fractional integro-differential operators of the FOPID controller. According to Caputo's definition, the $\alpha^{th}$ order derivative of a function $f_x(t)$ with respect to time is given by (1).

$$_0D_t^\alpha f_x(t) = \frac{1}{\Gamma(m-\alpha)} \int_0^t \frac{D^m f_x(t)}{(t-\tau)^{\alpha+1-m}} d\tau, \quad (1)$$
$$\alpha \in \Re^+, m \in Z^+, m-1 \leq \alpha < m$$

### B. Microgrid System and the Controller Structure

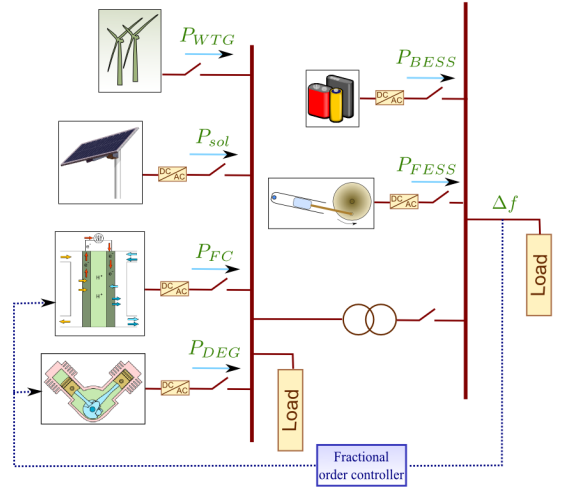

Fig. 1: Schematic of a microgrid with different connected energy sources.

Fig. 1 shows the schematic of the microgrid [7] used in the present study with various power generating units like the wind turbine, photovoltaic cell, fuel cells and diesel energy generator etc. There are also a battery and a flywheel energy storage system in the microgrid. The FOPID controller gives the control signal to the fuel cell (FC) and the diesel energy generator (DEG) based on the frequency deviation in the microgrid and tries to minimize the stochastic fluctuations in the grid frequency. Fig. 2 shows the block diagram schematic of the microgrid with the transfer functions of the individual components. The parameters of different components of the microgrid system are adopted from [7] as follows: $K_{WTG} = K_{FESS} = K_{BESS} = 1$, $D = 0.015$ pu/Hz, $2H = 0.1667$ pu s, $T_{FESS} = 0.1$ s, $T_{BESS} = 0.1$ s, $T_{FC} = 0.26$ s, $T_{WTG} = 1.5$ s, $T_g = 0.08$ s, $T_t = 0.4$ s, $T_{I/C} = 0.004$ s, $T_{IN} = 0.04$ s, $R = 3$ Hz/pu.

The FOPID controller is used to minimize the system frequency fluctuations in the microgrid and ensure better power quality. The transfer function representation of a FOPID controller is given by (2)

$$C(s) = K_p + K_i/s^\lambda + K_d s^\mu \quad (2)$$

This typical controller structure has five independent tuning knobs i.e. the three controller gains $\{K_p, K_i, K_d\}$ and two fractional order operators $\{\lambda, \mu\}$. For $\{\lambda, \mu\} = 1$ the controller structure (2) reduces to the classical PID controller in parallel structure. Few recent research results show that band-limited implementation of FOPID controllers using higher order rational transfer function approximation of the integro-differential operators gives satisfactory performance in industrial automation [25]. The Oustaloup's recursive approximation (ORA), which has been used to implement the integro-differential operators in frequency domain is given by (3), representing a higher order analog filter [10].

$$s^\alpha \approx K_f \prod_{k=-N}^{N} \frac{s + \omega'_k}{s + \omega_k} \quad (3)$$

where, the poles, zeros, and gain of the filter can be recursively evaluated as (4).

$$\omega_k = \omega_b \left(\frac{\omega_h}{\omega_b}\right)^{\frac{k+N+\frac{1}{2}(1+\alpha)}{2N+1}}, \omega'_k = \omega_b \left(\frac{\omega_h}{\omega_b}\right)^{\frac{k+N+\frac{1}{2}(1-\alpha)}{2N+1}}, K_f = \omega_h^\alpha \quad (4)$$

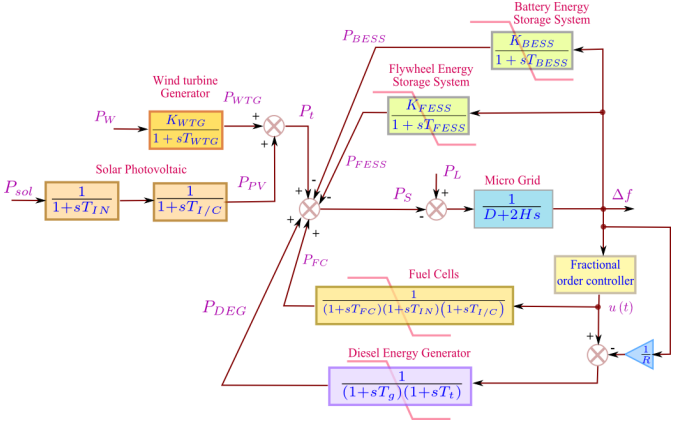

Fig. 2: Block diagram schematic of the microgrid used in the study.

The frequency deviation signal $\Delta f(t)$ is passed through the filter (3) and the output of the filter can be regarded as an approximation to the fractionally differentiated or integrated signal $D^\alpha [\Delta f(t)]$ which are then linearly combined with PID gains to obtain the final control signal for controller structure (2). In (3)-(4), $\alpha$ is the order of the differ-integration, $(2N+1)$ is the order of the analog filter and $(\omega_b, \omega_h)$ is the expected fitting range. Here, 5th order ORA is adopted for the FO operators within a chosen frequency band $\omega \in \{10^{-2}, 10^2\}$ rad/sec for the constant phase elements (CPEs) [10-12].

The controller output actuates the FC and DEG to control these units as they require expensive fuels. Depending on the grid frequency fluctuation, the controller sends a signal to the respective actuators i.e. like the hydrogen flow rate in the FC and mass flow rate of oil in DEG. Such a scheme can regulate the amount of power delivered by these devices to the microgrid to minimize the operating costs. The motor speed to run the flywheel and battery input current are directly taken from the grid frequency oscillation signal without the intervention of the controller as these devices does not need sophisticated control [7].

Here, the rated electrical specifications are adopted from [7] as the nominal case (without stochastic fluctuations), where all the powers are represented in pu with respect to the total base load demand of 410 kW (1 pu). The electrical load specifications of [7] are incorporated in the present model in the form of output saturation in each of the power producing components. The rated wind and solar powers are considered as 100 kW (0.244 pu) and 30 kW (0.073 pu) respectively with total renewable generation of 130 kW (0.317 pu). Similarly, the maximum power generated/absorbed by BESS and FESS are considered ±0.11 pu. In the case of the rapid power producing elements (FC/DEG), the lower bound of the saturation is zero since they cannot absorb power and the upper bound is 0.48 pu and 0.45 pu respectively. In [7], a sudden 0.1 pu (≈ 41 kW) of load disturbance has been introduced to study the performance of the control system. This has been improved in the present paper with a step-wise increase in the demand power having small stochastic fluctuations.

An upper and lower saturation limit is imposed on each energy storage element along with rate constraint nonlinearity to avoid any possible mechanical shock due to sudden large frequency fluctuation. The dynamical models in Fig. 1 represents small signal linearized transfer functions which captures the dynamic characteristics at a specific operating point [9]. The upper and lower limits of the output saturation will restrict the extraction/storage of large amount of power from/to a particular element than its rated values. The output saturations (in pu) and rate constraints for different elements are:

$|P_{FESS}| < 0.11, |P_{BESS}| < 0.11, 0 < P_{FC} < 0.48, 0 < P_{DEG} < 0.45$,
$|\dot{P}_{FESS}| < 0.05, |\dot{P}_{BESS}| < 0.05, |\dot{P}_{FC}| < 1, |\dot{P}_{DEG}| < 0.5$.

The microgrid has different control hierarchy like the local control, secondary control, centralized control etc. [26] and here, the secondary control scheme is adopted. In case of emergency where one subsystem needs to be disconnected, the centralized control loop overrides the functioning of the secondary control loop and disconnects the subsystem.

### C. Characteristic Changes of the Wind and Solar Power Generation and the Demand Load

Large deterministic drift and small stochastic power fluctuations [9] are considered for the wind generation, solar generation and load demand and are modeled in a general template here. This kind of template gives rise to a time-series with small stochastic fluctuations about the mean generated or demand power. In addition, the models take into account a sudden change in the mean value to represent real case scenarios where there are significant variations of these parameters. The general template for these is chosen as (5)

$$P = \left(\left(\phi\eta\sqrt{\beta}\left(1 - G(s)\right) + \beta\right)/\beta\right)\Gamma = \chi \cdot \Gamma \quad (5)$$

where, $P$ represents the power output of the solar, wind or the load model, $\phi$ is the stochastic component of the power, $\beta$ contributes to the mean value of the power, $G(s)$ is a low pass filter, $\eta$ is a constant in order to normalize the generated or demand power ($\chi$) to match the per unit (pu) level, $\Gamma$ is a time dependent switching signal with a gain which dictates the sudden fluctuation in mean value for the stochastic power output. Due to the sudden change in base value along with stochastic fluctuations, the source of such uncertain behavior in the power generation and demand can be modelled in a same template, having different parameters as studied in [9].

For the wind power generation the parameters of (5) are:
$\phi \sim U(-1,1)$, $\eta = 0.8$, $\beta = 10$, $G(s) = 1/(10^4 s + 1)$ and
$$\Gamma = 0.24 h(t) - 0.04 h(t-140) \quad (6)$$
where, $h(t)$ is the Heaviside step function.

For the solar power generation, the parameters of (5) are:
$\phi \sim U(-1,1)$, $\eta = 0.1$, $\beta = 10$, $G(s) = 1/(10^4 s + 1)$ and
$$\Gamma = 0.05 h(t) + 0.02 h(t-180) \quad (7)$$

For the demand load, the parameters of (5) are: $\phi \sim U(-1,1)$, $G(s) = (300/(300s+1)) + (1/(1800s+1))$, $\eta = 0.9$, $\beta = 10$ and

$$\Gamma = (1/\chi) \begin{bmatrix} 0.9 h(t) + 0.03 h(t-110) \\ +0.03 h(t-130) + 0.03 h(t-150) \\ -0.15 h(t-170) + 0.1 h(t-190) \end{bmatrix} + 0.02 h(t) \quad (8)$$

### D. Objective Function for Optimization

For effective functioning of the microgrid system, the controller gains and fractional integro-differential orders need to be tuned. For the controller design problem, the objective function in (9) is considered. It consists of the integral of two weighted terms, which try to minimize the frequency deviation in the microgrid ($\Delta f$), as well as the incremental control signal ($\Delta u$).

$$J = \int_{T_{min}=100}^{T_{max}=220} \left[ w(\Delta f)^2 + ((1-w)/K_n)(\Delta u)^2 \right] dt \quad (9)$$

where, $w$ dictates the relative importance of the two objectives (i.e. Integral of Squared Error – ISE and Integral of squared Deviation of Control Output - ISDCO) [15] and its value is taken as 0.7. $K_n = 10^4$ is the normalizing constant to scale ISE and ISDCO in uniform scale. For any fixed structure controller there is always a trade-off between the conflicting objectives of load disturbance rejection (reducing $\Delta f$ to zero very quickly) and the amount of control effort ($\Delta u$) required. Previous investigations [15] have shown that the problem is inherently multi-objective and to have a fast suppression of load disturbance, a higher amount of controller effort is required. The choice of $w$ as 0.7 in the present case indicates that the design gives more importance to the fast suppression of the microgrid frequency oscillations in comparison to the higher value of control signal.

### E. Kriging Based Global Optimization

Kriging models have been shown to be very expedient in accurate global approximations of the design space [27]. These models can approximate both linear and nonlinear trends in the design space. Their flexibility arises from the fact that different spatial correlation functions can be employed for building the approximation. Additionally, the kriging models can be built either to give more importance to the training dataset, by providing an exact interpolation, or they may be built to have a smooth inexact interpolation [28]. For the purpose of constructing a surrogate model, consider a set of $k$ design sites $S = [s_1 \cdots s_k]$ with $s_i \in \Re^n$ and corresponding model responses $Y = [y_1 \cdots y_k]^T$ with $y_i \in \Re^q$. The data is normalized to satisfy the following conditions in (10).

$$\begin{aligned} m[S_{:,j}] &= 0, \quad \sigma[S_{:,j}, S_{:,j}] = 1, \quad \forall j \in \{1,2,\cdots,n\} \\ m[Y_{:,j}] &= 0, \quad \sigma[Y_{:,j}, Y_{:,j}] = 1, \quad \forall j \in \{1,2,\cdots,q\} \end{aligned} \quad (10)$$

where $X_{:,j}$ is the vector represented by the $j^{th}$ column in matrix $X$, and $m[\cdot]$ and $\sigma[\cdot,\cdot]$ denote the mean and the covariance respectively.

A kriging model combines a global model with localized departures [29]. The model $\hat{y}$ that gives the deterministic response $y(x) \in \Re^q$, for an $n$ dimensional input $x \in \Gamma_z \subseteq \Re^n$ as a realization of a regression model $\Psi$ and a stochastic process,

$$\hat{y}_l(x) = \Psi(\zeta_{:,l}, x) + z_l(x), \quad l = 1, \cdots, q \quad (11)$$

The regression model $\Psi$ is taken as a linear combination of $p$ chosen functions $f_j : \Re^n \mapsto \Re, \forall j \in \{1,2,\cdots,p\}$ and can be expressed as

$$\begin{aligned} \Psi(\zeta_{:,l}, x) &= \zeta_{1,l} f_1(x) + \cdots + \zeta_{p,l} f_p(x) \\ &= [f_1(x) \cdots f_p(x)] \zeta_{:,l} = f(x)^T \zeta_{:,l} \end{aligned} \quad (12)$$

where, the coefficients $\{\zeta_{k,l}\}$ are the regression parameters. The random process $z$ is assumed to have zero mean and covariance between $z(w)$ and $z(x)$ is given by

$$E[z_l(w) z_l(x)] = \sigma_l^2 \Phi(\theta, w, x), \quad \forall l \in \{1,2,\cdots,q\} \quad (13)$$

where $\sigma_l^2$ is the process covariance for the $l^{th}$ component of the response and $\Phi(\theta, w, x)$ is the correlation model with parameters $\theta$.

For calculating the kriging predictor for the set $S$ of design sites, the design matrix $F \in \Re^{k \times p}$ with $F_{ij} = f_j(s_i)$ can be expressed as (14).

$$F = [f(s_1) \cdots f(s_k)]^T \quad (14)$$

where, $f(x)$ is defined in (12). Additionally the matrix $Q$ is defined to be that of the stochastic process correlations between $z$'s at the design sites,

$$Q_{ij} = \Phi(\theta, s_i, s_j) \quad \forall i, j \in \{1,2,\cdots,k\} \quad (15)$$

At a previously unsampled location $x$, let

$$q(x) = [\Phi(\theta, s_1, x) \cdots \Phi(\theta, s_k, x)]^T \quad (16)$$

represent the vector of correlations between $z$'s at the design sites and $x$.

For the regression problem of the form (17),

$$F\zeta \approx Y \quad (17)$$

the generalized least square solution (with respect to $Q$) is

$$\zeta^* = (F^T Q^{-1} F)^{-1} F^T Q^{-1} Y \quad (18)$$

and the corresponding kriging predictor can be expressed as

$$\begin{aligned} \hat{y}(x) &= q^T Q^{-1} Y + (F^T Q^{-1} q - f)^T \zeta^* \\ &= f(x)^T \zeta^* + q(x)^T Q^{-1} (Y - F\zeta^*) \end{aligned} \quad (19)$$

The mean squared error estimate $\varphi(x)$ of the predictor can be obtained as (20).

$$\varphi(x) = \sigma^2 \left(1 + b^T \left(F^T Q^{-1} F\right)^{-1} b - q^T Q^{-1} q\right) \quad (20)$$

where, $b = F^T Q^{-1} q - f$ and $\sigma^2$ is the maximum likelihood estimate of the variance.

The correlation models considered in this paper are of the form (21) [29].

$$\Phi(\theta, w, x) = \prod_{j=1}^{n} \Phi_j \left(\theta, w_j - x_j\right) \quad (21)$$

which are essentially products of stationary one dimensional correlations. Table 1 shows the different correlation functions used in the present study.

TABLE 1: DIFFERENT CORRELATION FUNCTIONS FOR KRIGING MODEL

| Name | $\Phi_j(\theta, d_j)$ where $d_j = w_j - x_j$ |
|---|---|
| Exponential | $e^{(-\theta_j \lvert d_j \rvert)}$ |
| Gaussian | $e^{(-\theta_j d_j^2)}$ |
| Linear | $\max\{0, 1 - \theta_j \lvert d_j \rvert\}$ |
| Spherical | $1 - 1.5\xi_j + 0.5\xi_j^3$, where $\xi_j = \min\{1, \theta_j \lvert d_j \rvert\}$ |
| Spline | $\psi(\xi_j) = \begin{cases} 1 - 15\xi_j^2 + 30\xi_j^3 & \text{for } 0 \le \xi_j \le 0.2 \\ 1.25(1 - \xi_j)^3 & \text{for } 0.2 < \xi_j < 1 \\ 0 & \text{for } \xi_j \ge 1 \end{cases}$ where $\xi_j = \theta_j \lvert d_j \rvert$ |

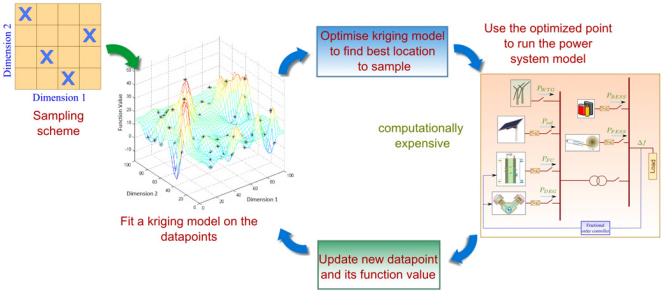

Fig. 3 Schematic of kriging based optimization method.

Fig. 3 shows a schematic of the kriging based surrogate modelling and optimization process. Initially a symmetric Latin hypercube sampling scheme is constructed in the design space of the fractional order controller (i.e. the three gains and the two fractional orders). This kind of a sampling scheme is a trade-off [30] between the simple Latin hypercube sampling (which might not cover the entire design space uniformly) and the computationally expensive 'space-filling' Latin hypercube sampling [31] methods. The bounds of the design variables $\{K_p, K_i, K_d, \lambda, \mu\}$ are kept between $\{0,0,0,0,0\}$ and $\{5,5,5,2,2\}$ respectively and 50 initial sampling points are generated using this scheme. Using these sampling points, the kriging model is constructed with one of the correlation functions as in Table 1. The candidate point approach [32] is used on the kriging model, for selecting the best location to sample in the next iteration. It works by creating two groups of candidates, one which is obtained by perturbing the best point obtained till the present iteration and the other which is generated by uniformly selecting points from the whole decision space. The kriging based response surface is used to predict the objective function values at the candidate points and is used to calculate the response surface criterion. The distance criterion is calculated which measures the distance of each candidate point to the existing set of sampled points. The weighted sum of these two criteria is used to determine the best candidate point. This best candidate point is then used to sample the computationally expensive objective function (by running the microgrid control model) and the actual value of the objective function is obtained. The kriging model is updated with this new value of the objective function and the process is iterated until a specified number of function evaluations are completed.

### III. SIMULATION AND RESULTS

#### A. Performance Evaluation of the Kriging Based Optimizers and Different Controller Structures

Since the objective function is stochastic in nature (i.e. the same value of controller parameters gives slightly different objective function values in each instance), the function is evaluated multiple times (10 in this case) and the expected value of the objective function is considered for optimization. Calling the objective function multiple times in this fashion is counted as one expensive function evaluation and the total number of these expensive function evaluations is limited to 150. The kriging based optimization is run with the different correlation models and the statistical results for 30 independent runs along with the best found expected minima (since the model contains stochastic components) of the objective (9) – $J_{min}$ are reported in Table 2. A comparison is done with a standard Genetic Algorithm (GA) which is run with 10 populations for 15 generations (so that the total number of function evaluations is 150) and the statistical results for the same are also reported in Table 2. The number of elite individuals is taken as 2 and the crossover and mutation fraction are taken as 0.8 and 0.2 respectively. The corresponding best found parameters for the PID and the FOPID controller are reported in Table 3. During the optimization four different controller structures are called depending on ranges of values for FOPID orders i.e. $\{\lambda, \mu\}$ greater and less than one, since the ORA can only approximate FO operators less than unity. For $\{\lambda, \mu\} > 1$ the nominal FO part is rationalized using ORA and an additional integrator/differentiator is added within the controller structure in series with the higher order rational approximation [10, 12].

The results in Table 2 indicate that the FOPID controller consistently outperforms the PID controller with all of the optimization algorithms. Also, the kriging based optimization algorithm with the spline correlation function gives the best result for both the PID/FOPID controllers. In both cases, the kriging based surrogate method outperforms the GA. Also the mean and standard deviation of the kriging based optimisation algorithms in Table 2 are very small as compared to the GA. This occurs since the GA is not able to find stable solutions in the many of the runs. Therefore the kriging based method not only gives better accuracy, but also consistently gives stable near optimal solutions every time. Table 2 also shows that the spline correlation model gives the best average solution for both PID/FOPID controller amongst the five correlation models in Table 1.

It is known in FO control that the integral order of FOPID higher than unity i.e. $\lambda>1$ makes the overall system faster, whereas $\lambda<1$ makes it slower [12]. On contrary the control effort for $\lambda>1$ increases drastically to ensure faster time response and $\lambda<1$ produces smaller control effort. Since the cost function (9) has got two parts balancing the impact of fast tracking and control effort as a weighted sum, under different circumstances integro-differential orders may be less than or greater than one, although in Table 3 both the optimum FO orders are <1.

TABLE 2: STATISTICAL RESULTS OF 30 INDEPENDENT RUNS OF DIFFERENT KRIGING BASED OPTIMIZATION WITH PID/FOPID CONTROLLER

| Kriging model /optimizer | Controller | Statistics of $J$ for 30 independent runs | | |
|---|---|---|---|---|
| | | $J_{min}$ | Mean | Standard deviation |
| Exponential | PID | 0.01413 | 0.01830 | 0.00523 |
| | FOPID | 0.00396 | 0.01055 | 0.00758 |
| Gaussian | PID | 0.01407 | 0.01655 | 0.00251 |
| | FOPID | 0.00391 | 0.00977 | 0.01473 |
| Linear | PID | 0.01434 | 0.01786 | 0.00625 |
| | FOPID | 0.00404 | 0.00994 | 0.00773 |
| Sphere | PID | 0.01406 | 0.01807 | 0.00553 |
| | FOPID | 0.00401 | 0.00892 | 0.00522 |
| Spline | PID | *0.01392* | *0.01639* | 0.00272 |
| | FOPID | *0.00382* | *0.00604* | 0.00230 |
| GA | PID | 0.01419 | 6.95936 | 27.77352 |
| | FOPID | 0.00421 | 2.92570 | 11.21648 |

TABLE 3: BEST PID/FOPID CONTROLLER PARAMETERS

| Kriging model /optimizer | Controller | Best controller parameters | | | | |
|---|---|---|---|---|---|---|
| | | $K_p$ | $K_i$ | $K_d$ | $\lambda$ | $\mu$ |
| Exponential | PID | 3.613 | 1.822 | 0.344 | - | - |
| | FOPID | 0.984 | 3.359 | 1.426 | 0.677 | 0.623 |
| Gaussian | PID | 3.666 | 1.903 | 0.333 | - | - |
| | FOPID | 2.461 | 5.000 | 0.948 | 0.926 | 0.744 |
| Linear | PID | 4.150 | 1.250 | 0.350 | - | - |
| | FOPID | 2.204 | 3.155 | 1.233 | 0.768 | 0.705 |
| Sphere | PID | 3.678 | 1.351 | 0.342 | - | - |
| | FOPID | 2.450 | 4.750 | 0.950 | 0.860 | 0.780 |
| Spline | PID | 3.712 | 1.391 | 0.333 | - | - |
| | FOPID | 0.950 | 4.350 | 1.250 | 0.660 | 0.700 |
| GA | PID | 3.124 | 1.087 | 0.324 | - | - |
| | FOPID | 1.703 | 2.166 | 1.310 | 0.992 | 0.654 |

Fig. 4 shows a single realization of the stochastic process model for power generation by the WTG, PV and the load (5) which is used in the simulation study. It is observed that there are significant fluctuations which would be reflected in the system frequency and the controller needs to take appropriate action to damp out these oscillations considering all possible realization of the stochastic objective function (9), thereby locating an expected minima [33, 34] based on multiple runs of the same model. Therefore, the role of the kriging model used here is to approximate a dynamically evolving noisy function taking into account the correlations among the different controller parameters to find out the expected global minima in less number of iterations. Such reduction in number of iterations would facilitate the online implementation of tuning such controllers in future [7]. Significant non-stationary nature of the stochastic fluctuation with drift in renewable generation and demand load are evident in Fig. 4. In the present design framework, it is considered that the microgrid was operating at 1 pu load during $0<t<100$ sec and the control system performance has been evaluated then for a finite time horizon of $100<t<220$ sec considering change in both the demand load and renewable generations (Fig. 4).

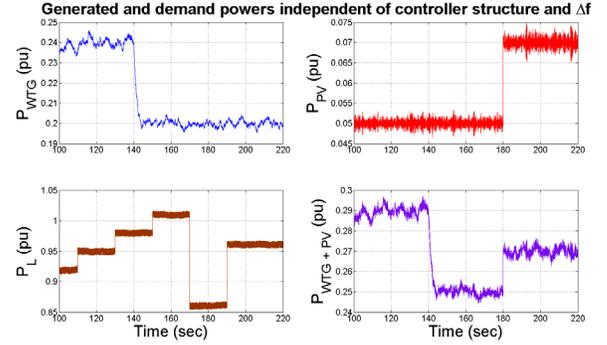

Fig. 4 One realization of the stochastic generated and demand powers independent of the controller structure.

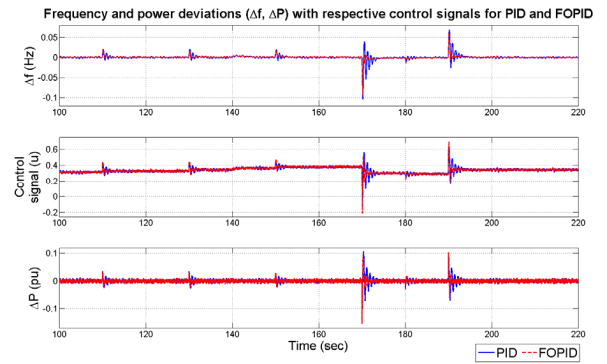

Fig. 5 Frequency deviation, control signal and power deviation of the microgrid with best obtained PID/FOPID controllers.

The microgrid frequency deviation ($\Delta f$), control signal ($u$) and the deficit/excess power ($\Delta P$) corresponding to the best obtained PID and the FOPID controller are shown in Fig. 5. It is evident that the FOPID controller outperforms the PID controller since it results in less frequency fluctuation $\Delta f$ (hence better power quality), faster damping of the deficit grid power $\Delta P$ with less control signal (hence less actuator size requirements for the FC and DEG).

The corresponding powers produced by the FESS, BESS, FC and DEG are reported in Fig. 6. It can be observed that there are less power fluctuations with the FOPID controller than the PID controller. This implies that if the FOPID controller is used, the sizing of these energy supply and storage systems can be made smaller. Also there are less requirements of supplying and storing power to supress the microgrid frequency fluctuations. This makes the overall system more energy efficient.

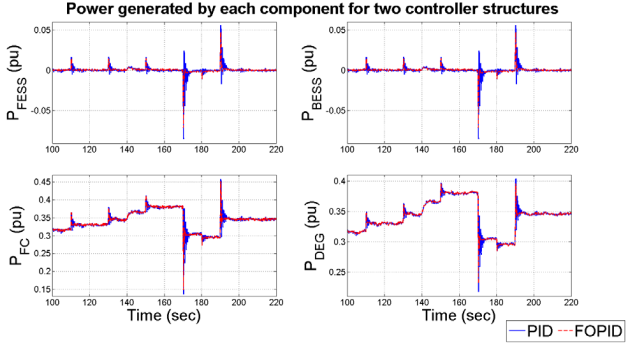

Fig. 6 Individual powers of the different components of the microgrid with the best obtained PID/FOPID controllers.

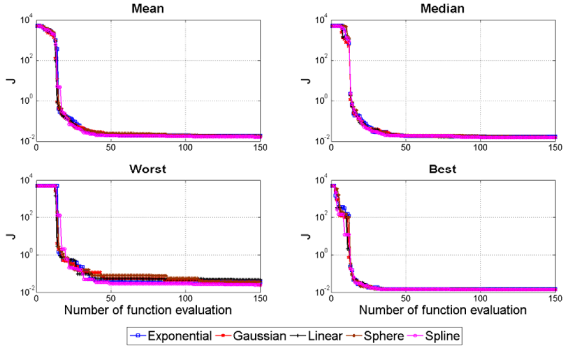

Fig. 7 Convergence of five kriging methods for PID controller.

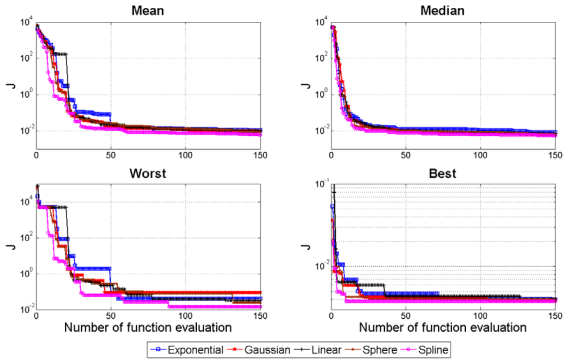

Fig. 8 Convergence of five kriging methods for FOPID controller.

Fig. 7 and Fig. 8 show four standard statistical measures (mean, median, best and worst case) of the convergence curves in semi-log scale, for 30 independent runs of the kriging based optimization employing different correlation functions for the PID and the FOPID controller respectively. The corresponding curves for the GA are plotted in Fig. 9. It is clear that to find the best solutions, the GA takes more number of iterations whereas all the kriging based optimizers converges to the best solutions very quickly. Therefore, the kriging solutions would significantly outperform the GA even if less number of function evaluations (than 150) are used. Also, if there are significant stochastic deviations in the objective function, then calling the function only 10 times (as done in this case) would not be suitable for approximating an expected value and more number of samples need to be taken. In such cases, the kriging would significantly outperform the GA as well. As also evident from Fig. 7-Fig. 8 that for both PID and FOPID controller the spline correlation function is capable of locating lower value of the objective functions in all the four cases of statistical measures which is in agreement with the findings reported in Table 2.

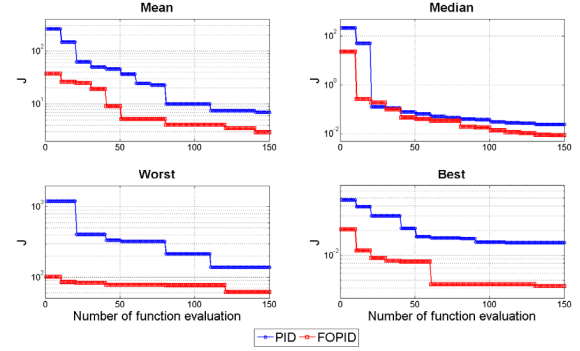

Fig. 9 Convergence of GA based tuning of PID/FOPID controllers.

### B. Parametric Robustness of the Optimum Solutions

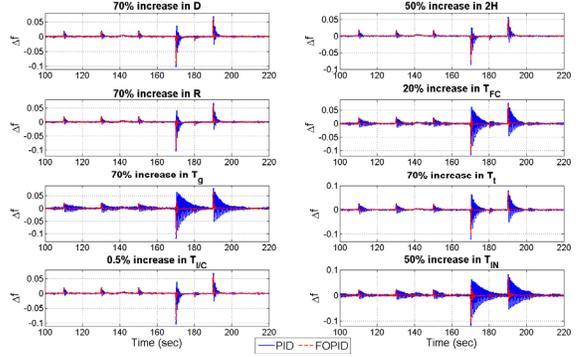

Fig. 10 Robustness for increase in system parameters.

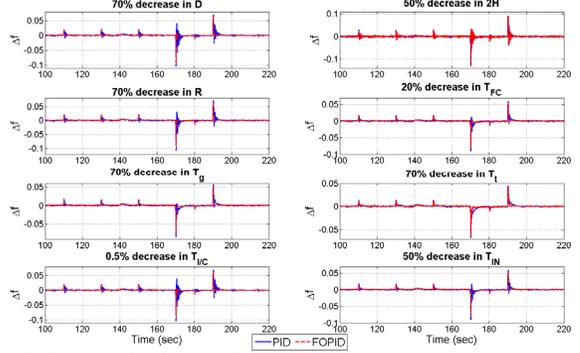

Fig. 11 Robustness for decrease in system parameters.

The best solutions for the PID/FOPID controllers are now obtained from Table 2 and Table 3 (as the nominal case) and their robustness is tested for perturbed case of different microgrid parameters. The expected values of the objective function (9) are noted in Table 4 considering multiple runs of the stochastic model. The corresponding microgrid frequency deviation for both the case of parameter increase and decrease (in $D$, $2H$, $R$, $T_{FC}$, $T_g$, $T_t$, $T_{I/C}$, $T_{IN}$) are shown in Fig. 10 and Fig. 11 respectively. It can be observed from Table 4 and also Fig. 10 and Fig. 11 that the PID controller is less robust than the FOPID controller in terms of increased value of $J$ under perturbed condition and increase in $\Delta f$. This establishes the superiority of using FOPID controller in such a microgrid frequency control problem with large positive/negative uncertainty in the system parameters. It is also evident from Fig. 10-Fig. 11 that the increase in system parameters is much detrimental than a decrease in equal amount in most cases, although the ill-effects are significantly smaller with an FOPID. Among different parameters the interconnection device, FC and inverter time constants ($T_{I/C}$, $T_{FC}$, $T_{IN}$) are found to be gradually most susceptible ones which cause performance deterioration much faster than perturbing other microgrid parameters.

TABLE 4: ROBUSTNESS FOR **PERTURBATION IN** MICROGRID PARAMETER

| Microgrid parameters | Perturbation (±) | Controller | $J$ for increase in parameter | $J$ for decrease in parameter |
|---|---|---|---|---|
| $D$ | 70% | PID | 0.01381 | 0.01403 |
|  |  | FOPID | 0.003802 | 0.003847 |
| $2H$ | 50% | PID | 0.007543 | 0.04399 |
|  |  | FOPID | 0.002453 | 0.0147 |
| $R$ | 70% | PID | 0.01397 | 0.01378 |
|  |  | FOPID | 0.00381 | 0.00397 |
| $T_{FC}$ | 20% | PID | 0.02079 | 0.01222 |
|  |  | FOPID | 0.00443 | 0.003411 |
| $T_g$ | 70% | PID | 0.02543 | 0.01207 |
|  |  | FOPID | 0.004496 | 0.003242 |
| $T_t$ | 70% | PID | 0.01928 | 0.01134 |
|  |  | FOPID | 0.004688 | 0.003126 |
| $T_{I/C}$ | 0.5% | PID | 0.01364 | 0.01512 |
|  |  | FOPID | 0.003776 | 0.004133 |
| $T_{IN}$ | 50% | PID | 0.02675 | 0.01424 |
|  |  | FOPID | 0.004816 | 0.003722 |

### C. Effect of Actuator On-Off Switching Logic

Next, it is assumed that the FC and DEG does not supply any power if the grid frequency deviation is within a specified limit of $|\Delta f| < 0.05$ and only turns on when the FESS and BESS cannot supply/absorb enough power within a short period to damp the frequency deviation. It turns out that the scheme suffers from *chattering* which is common for such switched systems and sliding mode control. Chattering is detrimental for different components of the microgrid and hence is not desirable. To alleviate this issue to some extent a modification of the scheme is done such that it remains on for at least a minimum of 10 sec until the frequency deviations are within acceptable limits. However as soon as this happens and both the DEG and FC are cut out of the microgrid, the system frequency starts to deviate and the scheme again triggers the DEG and FC on, after a few milliseconds. Therefore, there are significant switching transients after almost every 10 seconds as shown in Fig 12 which is undesirable. The actuator on-off switching logic is implemented in Stateflow while modelling of the microgrid with FOPID is done using Simulink which is invoked by the optimization procedure coded in Matlab scripts. The whole model (with system nonlinearity in the form of rate constraint and output saturation, FOPID controller, stochastic forcing with jumps) is numerically integrated with the 3$^{rd}$ order accurate Bogacki-Shampine formula with a fixed step-size of 0.01 sec.

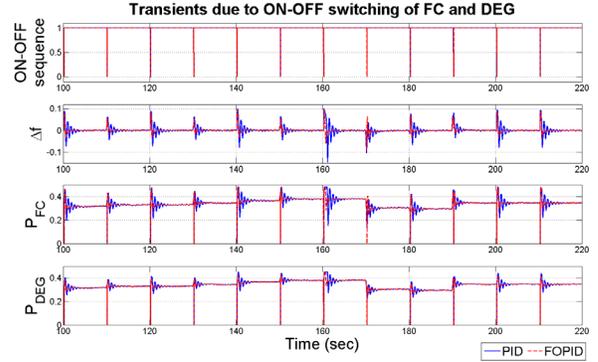

Fig. 12 Transients due to on-off switching of the actuators.

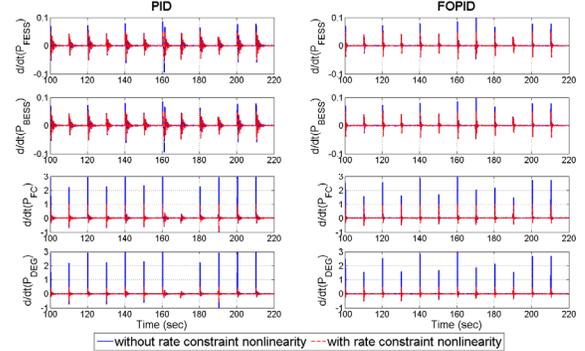

Fig. 13 Significant nonlinear operation of the energy storage/supplying elements with rate constraint nonlinearity.

Operation in significant nonlinear zone has also been shown in Fig. 13 along with the actuator on-off logic in Fig. 12 for the present microgrid system with PID/FOPID controller. It is evident from Fig. 13 that the rate of change of power in each element above or below the respective thresholds have been cut-off and a constant rate is maintained. Also the DEG and FC in Fig. 12 produces a constant amount of power once they reach their maximum limits due to the presence of the output nonlinearity in the system.

## IV. CONCLUSION

The paper proposes the use of fractional order controller for supressing the system frequency deviation in a nonlinear and stochastic model of a microgrid. Simulation results show that the FOPID controller is better than the standard PID controller under nominal operating condition and gives better robustness for large parametric uncertainty of the microgrid. A kriging based surrogate modelling and optimization technique is proposed which reduces the time taken for optimizing the controller parameters for the microgrid system. This can facilitate the online tuning of such controllers in future. Simulation results show that the kriging based optimization outperforms the standard genetic algorithm in terms of the quality of solutions and faster convergence.